# Pre-equilibrium α-particle emission as a probe to study α-clustering in nuclei

O.V. Fotina[1,2a], S.A. Goncharov[1,2], D.O. Eremenko[1,2], S.Yu. Platonov[1,2], O.A. Yuminov[2], V.L. Kravchuk[3], F. Gramegna[4], T. Marchi[4], M. Cinausero[4], M. D'Agostino[5], M. Bruno[5], G. Baiocco[5], L. Morelli[5], M. Degerlier[6], G. Casini[7], S. Barlini[7], S. Valdrè[7], S. Piantelli[7], G. Pasquali[7], A. Bracco[8], F. Camera[8], O. Wieland[8], G. Benzoni[8], N. Blasi[8], A. Giaz[8], A. Corsi[8], D. Fabris[9]
for the NUCL-ex and HECTOR collaboration

[1]Physical Department, Lomonosov Moscow State University, Moscow, Russia
[2]Skobeltsyn Institute of Nuclear Physics, Lomonosov Moscow State University, Moscow, Russia
[3]National Research Center "Kurchatov Institute", Moscow, Russia.
[4]INFN, Laboratori Nazionali di Legnaro, I-35020, Legnaro, Italy
[5]Dipartimento di Fisica, University of Bologna and INFN sezione di Bologna, Italy
[6]University of Nevsehir, Science and Art Faculty, Physics Department, Nevsehir, Turkey
[7]Dipartimento di Fisica, University of Firenze and INFN Sezione di Firenze, Italy
[8]Dipartimento di Fisica, University of Milano and INFN Sezione di Milano, Italy
[9] INFN Sezione di Padova, Italy

**Abstract.** A theoretical approach was developed to describe secondary particle emission in heavy ion collisions, with special regards to pre-equilibrium α-particle production. Griffin's model [1] of non-equilibrium processes is used to account for the first stage of the compound system formation, while a Monte Carlo statistical approach was used to describe the further decay from a hot source at thermal equilibrium. The probabilities of neutron, proton and α-particle emission have been evaluated for both the equilibrium and pre-equilibrium stages of the process. Fission and γ-ray emission competition were also considered after equilibration. Effects due the possible cluster structure of the projectile which has been excited during the collisions have been experimentally evidenced studying the double differential cross sections of p and α-particles emitted in the E=250MeV $^{16}$O +$^{116}$Sn reaction. Calculations within the present model with different clusterization probabilities have been compared to the experimental data.

## 1. Introduction

Double-differential spectra $\delta^2\sigma/\delta\Omega\delta E$ of emitted light charged particles have been measured in the reaction $^{16}$O+$^{116}$Sn at energies of 130MeV and 250MeV with the GARFIELD+HECTOR coupled set-ups at Laboratori Nazionali di Legnaro. The experimental distributions have been compared to our model predictions. A satisfactory agreement is observed for the proton distributions at both energies over the measured angular range as it can be seen in Figure 1 and Figure 2. On the contrary for α particles, while a quite good agreement persists between theory and experiment at large angles, a significant difference can be observed at forward directions at both energies as shown in Figure 3 and Figure 4. This effect is a result of the production of the secondary alpha particles during non-equilibrium stage of fusion nuclear reaction.

---

[a]Corresponding author: fotina@srd.sinp.msu.ru

A modified version of the statistical code PACE was used as a basis to describe the emission mechanisms: the main modification was related to the insertion of a non-equilibrium stage in the fusion reaction. The relaxation processes in the nuclear system occurring during the fusion reaction was accounted for by the exciton model. One of the more intricate questions is the description of the angular distributions of secondary particles emitted in the non-equilibrium stage of the reaction. In the present work we used two methods to describe the angular distributions. The first one (case *A*) is based on the exciton model of Griffin [1] modelling the orbital angular momentum of the emitted particle in the framework of the optical model. The second one (case *B*) is the hybrid exciton model suggested in [2, 3], where cluster/light ion induced reactions are considered. More detailed description of the two methods can be found in [7] and ref. therein. The best agreement to the data obtained by tuning the free parameters for case *B* is shown in Fig. 1-4.

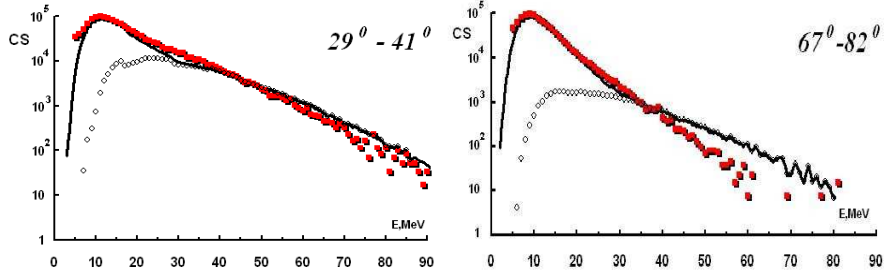

**Figure 1.** Double differential spectra (Cross Section (CS) in arbitrary units) for protons for the 250MeV $^{16}$O + $^{116}$Sn reaction. Experimental data are shown in red. Open circles show the pre-equilibrium part of the spectra. The continuous line is the sum of the predicted evaporative and pre-equilibrium contributions [4-6].

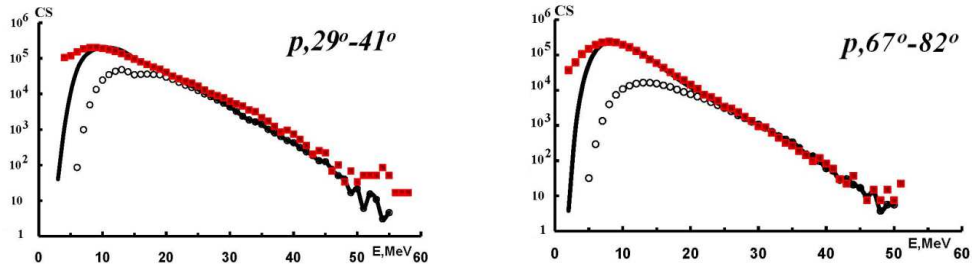

**Figure 2.** Double differential spectra (CS in arbitrary units) for protons for the 130MeV $^{16}$O + $^{116}$Sn reaction. Symbols are the same as in Fig. 1.

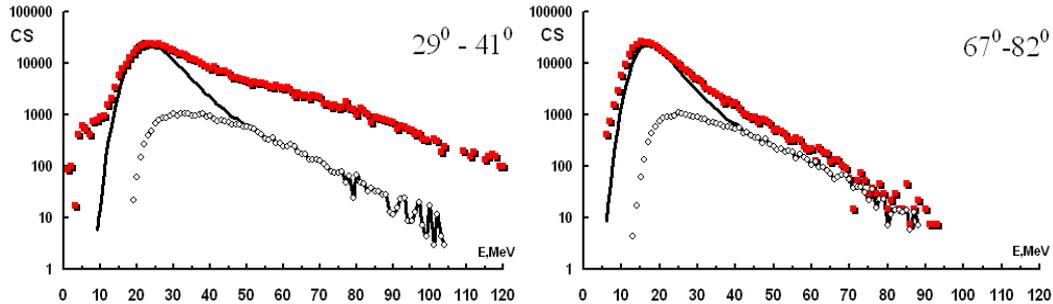

**Figure 3.** Double differential spectra (CS in arbitrary units) for α particles for the 250 Mev$^{16}$O + $^{116}$Sn reaction. Symbols are the same as in Figure 1.

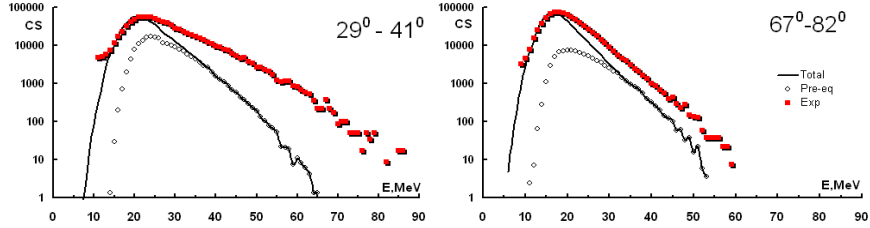

**Figure 4.** Double differential spectra (CS in arbitrary units) for α particles for the 130 Mev $^{16}$O + $^{116}$Sn reaction. Symbols are the same as in Figure 1.

Analysis of these results led to the conclusion that the pre-equilibrium emission contribution, whatever the implemented recipe, is not sufficient to explain the increased production of α-particles especially those observed in the experiment at forward angles. Additional effects, as the clustering structure of the projectile nucleus, must be considered. To verify this assumption we propose an approach to assess possible influences of the cluster structure of projectile in the calculations of the particle spectra.

## 2. Brief Description of the model.

A schematic description will be given on the first steps necessary to simulate the effect of the cluster structure of the projectile: in both cases, *A (Griffin exciton model)* and *B (Hybrid exciton model)*, the main parameter to be set is the initial number of excitons. According to the estimate from the work of E. Betak [8], the initial number of excitons for the $^{16}$O cases discussed above is $n_0=16$. The method for estimating the formation probability of the cluster, together with the algorithm used in the Monte-Carlo procedure to determine the exciton energies for cluster/light ion induced reactions, were presented in the work of M. Blann and M.B. Chadwick [9].

We used the expression $e = E\left(1 - (1-x)^{1/(n_0-1)}\right)$ in order to estimate the energy $e$ of the cluster consisting of $n_0$ initial excitons. Here $E$ is the initial projectile energy and $x$ is a random number with uniform distribution in (0,1). In our case, according to the energy conservation law, if $e_4$ is the α-cluster energy emitted by the oxygen projectile, the remaining carbon nuclei should have energy $e_{12}= E_{16O}-B_\alpha-e_4$. An example of the cluster energy distribution for $E_{16O}=250$ MeV is presented in Fig 5.

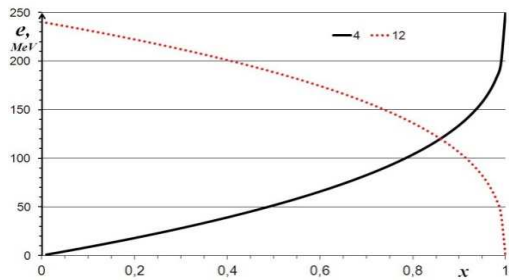

**Figure 5.** Energy distribution of α-clusters (black line $e_4$) and carbon (red dots $e_{12}$) from random number $x$.

As a first step we use the probability of the α-clusters formation as a free parameter.

## 3 Results and conclusions

The results of the calculated double differential cross sections for α-particles with different probabilities of α-clustering pre-formation in comparison with estimations ones without clustering (Case *A*) are shown in Figure 6. As it can be seen the inclusion of some pre-formed α-structure better describes the experimental data. Nevertheless these results open up some more specific questions: what are the real energy and angular distributions of this fast emitted particles and how can be the formation probability of cluster determined?

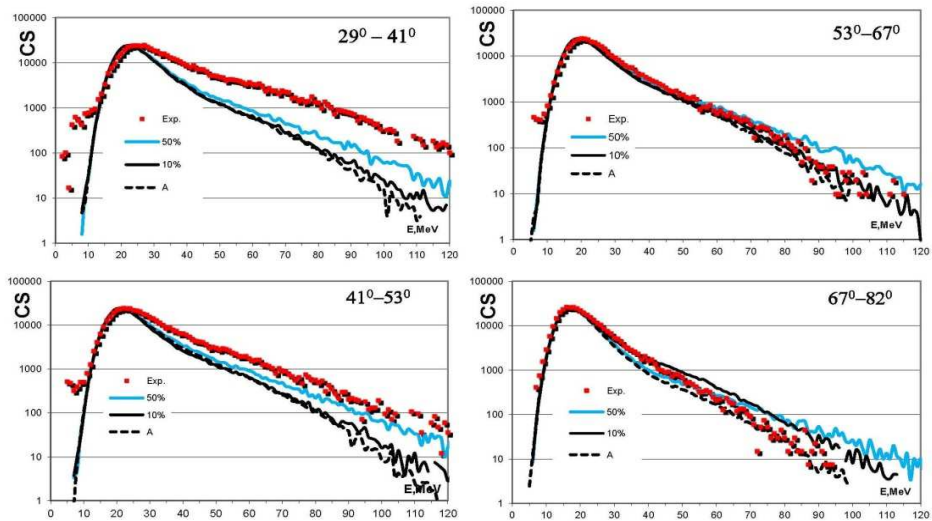

**Figure 6.** Double differential spectra (Cross-Section (CS) in arbitrary units) for α particles for the 250 MeV $^{16}$O + $^{116}$Sn reaction. Experimental data are shown in red. Calculated events in the case of different α-clustering pre-formation probabilities (10% black line and 50% cyan line) in the $^{16}$O projectile. The black dashed line is the case *A* estimations without of α-clustering in the Oxygen.

In order to answer to some of these questions in a model-independent way, we recently performed at the Legnaro National Laboratory an experiment aimed at investigating the α-particle emission from the hot $^{81}$Rb nucleus. The same compound nucleus has been formed at the same projectile velocity (16 AMeV) reactions with an α-cluster $^{16}$O projectile on $^{65}$Cu target and with a non-α-cluster $^{19}$F projectile on $^{62}$Ni target. The experiment was performed using the GARFIELD and RCo detection arrays. The main goal of this experiment is to measure and to compare the experimental pre-equilibrium α-particle yields for the two systems. It is expected, in fact, that any difference in the experimental results can be interpreted as a model-independent way to establish the influence of the projectile α-cluster structure on the spectra of α-particles emitted during the reaction phases. Experimental results will be even compared to model calculations as described in this contribution.